Dataset in Focus

# Datasets of Fire and Crime Incidents in Pampanga, Philippines

John Paul P. Miranda
College of Computing Studies, Mexico Campus, Don Honorio Ventura State University, Philippines
jppmiranda@dhvsu.edu.ph
(corresponding author)

Julieta M. Umali
College of Computing Studies, Bacolor Campus, Don Honorio Ventura State University, Philippines
jmumali@dhvsu.edu.ph

Aileen P. de Leon
College of Computing Studies, Bacolor Campus, Don Honorio Ventura State University, Philippines
apdeleon@dhvsu.edu.ph

**Abstract**

*Purpose* – The fire and crime incident datasets were requested and collected from two Philippine regional agencies (i.e., the Bureau of Fire Protection and the Philippine National Police). The datasets were used to initially analyze and map both fire and crime incidents within the province of Pampanga for a specific time frame.

*Method* – Several data preparation, normalization, and data cleaning steps were implemented to properly map and identify patterns within the datasets.

*Results* – The initial results also indicate the leading causes of fire and crimes are rubbish and acts against property. Fires mostly occur during the dry season in the province. Crime is particularly high during December, and most of the fire and crime incidents occur during the time when people are most active.

*Conclusion* - The dataset was able to present the temporal characteristics of the fire and crime incidents that occurred in the province of Pampanga.

*Recommendations* - Merge the existing dataset with the other datasets from other related agencies to get a bigger picture and produce more objective results that could be used for decision-making.

*Keywords* – dataset, fire, crime, incidents, rubbish, crimes against property, temporal

## VALUE OF THE DATASETS

The datasets were initially sought to compare the results from other studies (e.g., Bringula & Balahadia, 2019). Only the datasets for which the authors are currently residing at the time of the study were obtained. The fire dataset was requested and collected from the Philippine Bureau of Fire Protection (BFP) in Regional Office III located in the City of San Fernando, Pampanga. The crime dataset came from the Philippine National Police (PNP) Regional Office located in the same city. Prior requests were submitted to the Freedom of Information (eFOI) website of the Philippine government. The aforementioned process fast-tracked the release of the datasets. The datasets were completed and collected in 2020, before the implementation of lockdowns due to COVID-19 in the Philippines. The fire dataset contained useful data from 2013–2020. while the crime dataset contained data from 2017–2020. Previous years either had too much missing data or were mostly unreadable and useless.

Fire incidents are catastrophic in nature and bring havoc to a community (e.g., Oñate, 2022). The high volume of crimes tends to harm a community's development and the overall welfare of its residents. This is where the datasets may be useful, particularly for conducting computational tasks and achieving certain tasks that might be beneficial to the community as a whole (i.e., determining the leading causes of fire and predicting the occurrences of crimes). For example, the fire dataset can be used to map and find certain patterns of fire cases (Balahadia et al., 2020). Crime datasets could be used to determine the susceptibility of one location to becoming a crime hotspot as well as predict the probability of a specific crime occurring (Villarica et al., 2022). In addition to this, understanding and analyzing this data might be useful for crime mitigation and prevention (Bradley & Waller, 2017). The data is relevant to improving the systems within the city. The analysis of the two datasets is critical for making the best use of limited resources (i.e., equipment and manpower).

## DATASETS ACQUISITION

The fire and crime datasets have seven variables. Table 1 shows the fire dataset, which includes the date and time, location, and type of establishment involved, as well as identifiable names of people; the highest alarm level recorded; potential and identified causes and origins of fire incidents; and the estimated damage and whether the case is pending or closed.

For the crime dataset, the provincial name where the data is primarily stored is included. The station where the specific barangay where the crime occurred is also part of the variables. The recorded date and time of the crime, including the stages of the felony (i.e., acts and omissions punishable by law in the Philippines) and the actual violated laws, were also included.

Table 1. Description of the Variables

| Variable | Description | Data Type | Sample |
|---|---|---|---|
| **Fire Dataset** | | | |
| Date and time | Data and time of fire incident (PH time) | String | 01 0020H Jan 2014 |
| Location | The specific location of the fire incident | String | Sta. Lucia, Sitio Paroba, City of San Fernando |
| Involve establishment, name of owner/s and occupant/s | Type of establishment including the name of owners and occupants if any. | String | Residential [name redacted] |
| Alarm status | The highest alarm level of the fire incident recorded | String | 1st Alarm |
| Cause/ origin of fire and nature of the case | Determined cause and nature of fire | String | Flying lantern |
| Name of fatality/ injured | Name of the person who died or was injured due to the fire including the level of burn received by the person | Mixed (i.e., String, Number) | Negative |
| Status of case/ amount of damage | Estimated damage due to fire | Number | 10,000.00 |

Table 1. Description of the Variables (continuation)

| Variable | Description | Data Type | Sample |
|---|---|---|---|
| **Crime Dataset** | | | |
| PPO | Provincial office name | String | Pampanga PPO |
| Station | Municipal or city station where the crime is recorded | String | San Fernando |
| Barangay | Specific barangay name where the crime is committed | String | Dolores |
| Date committed | The recorded date of crime | Date | 7/7/2018 |
| Time committed | The recorded time of the crime | Time | 9:30 PM |
| Stages of felony | Recorded type of execution | String | RECKLESS IMPRUDENCE |
| Offense | Violated code/ law | String | MALICIOUS MISCHIEF |

Figure 1 and Figure 2 show a partial glimpse of what the dataset looked like in the prior analysis. The name and vehicular plate numbers are included. Due to the privacy laws in the Philippines, these were omitted, as seen in Figure 1. All other data that could identify a person was removed as part of the data cleaning process. During the data cleaning process, the data and time were separated from one another. The municipality/city and barangay names are separated from each other. The causes and nature of fire are also created as new variables. Several new variables were created during the data preparation of fire data. For example, the owner, occupants, fatalities, and injured count are all created separately. Another set of variables with the same names was created but contained only Boolean values (i.e., True and False).

*Figure 1.* A glimpse of the fire dataset in spreadsheet format

*Figure 2.* A glimpse of crime dataset in spreadsheet format

## CURRENT APPLICATION AND LIMITATIONS OF THE DATASETS

The fire dataset was used to map and observe any potential patterns that could help revisit the fire prevention programs. The fire dataset was analyzed based on its temporal attributes (De Leon & Miranda, 2022). On the other hand, the temporal attributes of the crime dataset were also analyzed to check for any potential patterns. The crime dataset is undergoing further analysis by combining it with the weather dataset requested by the Philippine Atmospheric, Geophysical, and Astronomical Services Administration (PAGASA) at the time of the writing of this study. Machine learning algorithms are also being tested to predict potential occurrences of a specific crime at a specific time and place.

## INITIAL RESULTS FROM DATA ANALYSES

### *Annual Recorded Fire and Crime Incidents*

Figure 3 illustrates the annual frequency of fire and crime in Pampanga. The number of recorded fire incidents in the province is relatively higher despite the implementation of a series of lockdowns in the country in 2020. Meanwhile, the total number of crimes recorded plummeted in the same year. The annual average of fire incidents from 2013 to 2020 was 7,111.5. On the other hand, the crime average from 2017 to 2020 was 380. When the two datasets are normalized and compared to one another from 2017 to 2020, their averages are 2.59 for fire and 3.85 for crime incidents.

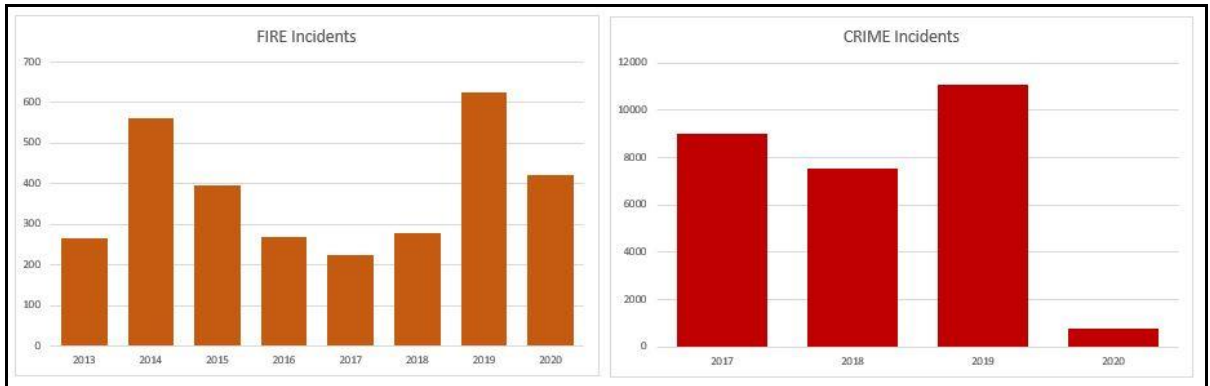

*Figure 3.* Annual frequency of fires (right) and crimes (left)

## *Monthly Recorded Incidents*

The monthly incidents of fire and crimes are illustrated in Figure 4. It shows that fire incidents start to climb in the month of October and peak in March. In the Philippines, fire prevention month is celebrated in March of every year. Fire incidents are relatively lower during the wet (i.e., rainy) season in the country. On the other hand, the crimes are higher during the month of December, as seen in Figure 4. When the fire and dataset are compared to one another after normalization, their averages are 2.40 for the fire and 3.37 for the crime.

## *Normalized Hourly Incidents*

Figure 5 illustrates the number of fires and crimes per hour. The data is normalized for better visualization. Reported fire incidents start to rise at 8:00 in the morning and fall after 6:00 in the evening. Meanwhile, crimes are rising from 6:00 AM to 11:00 PM. Based on these, the initial findings showed that fire and crimes are rising when people are more active (Bringula & Balahadia, 2019). The peak for fires and crimes happened between 2: 00–2: 59 and 5:00–5: 59 PM, respectively. Crime occurrences in the province are usually clustered within similar ranges. These findings supported earlier findings by Barrera et al. (2013).

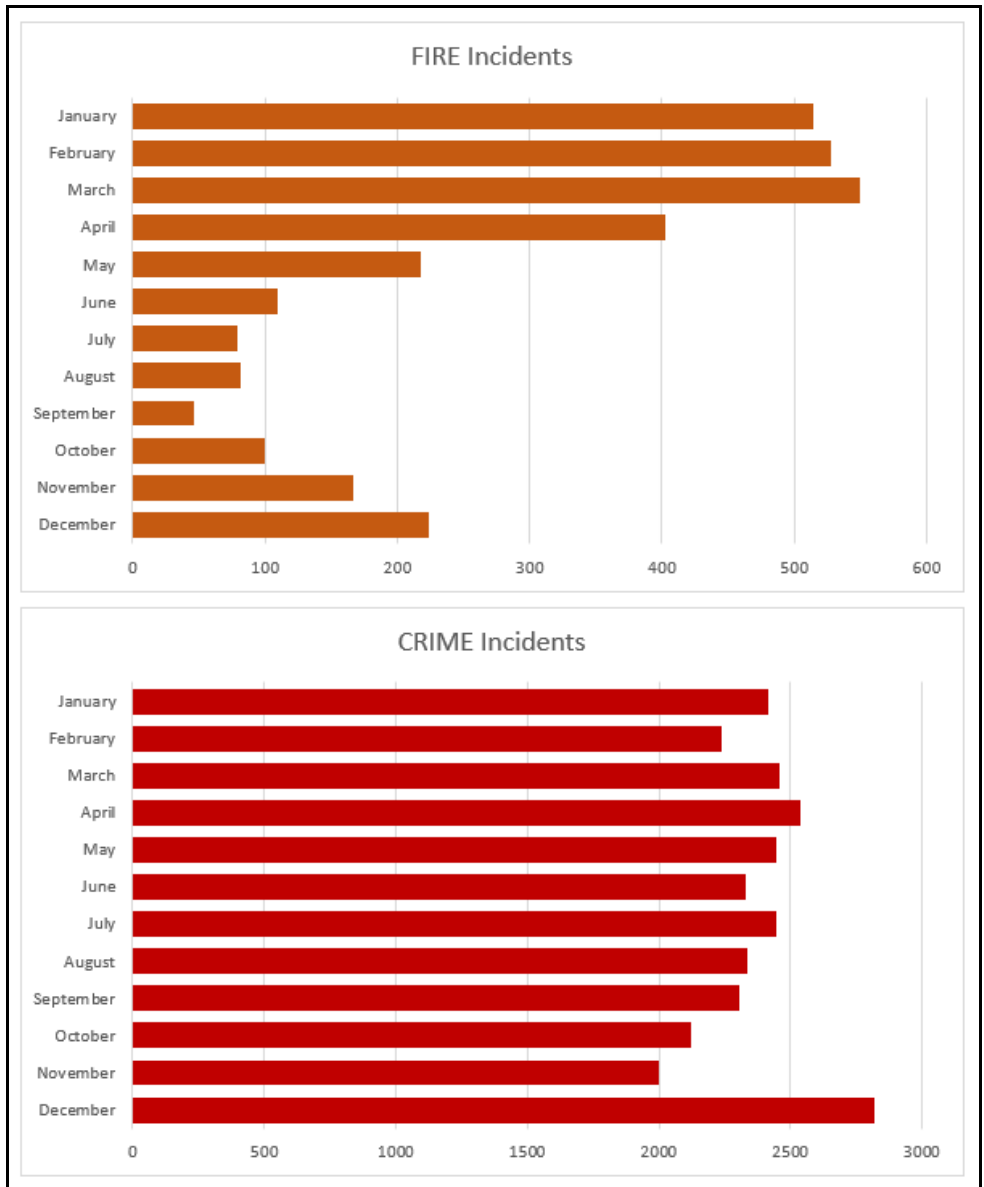

*Figure 4.* Monthly frequency of fires (top) and crimes (bottom)

## *Leading Causes*

Rubbish (n = 1,860) was found to be the leading cause and origin of fire incidents in the province. This is followed by faulty electrical connections (n = 572). This observation of faulty electrical connections is similar to the findings of Bringula & Balahadia in 2019. The incidents mostly happened in open green spaces such as grass fields (n = 1,708), followed by residential areas (n = 720).

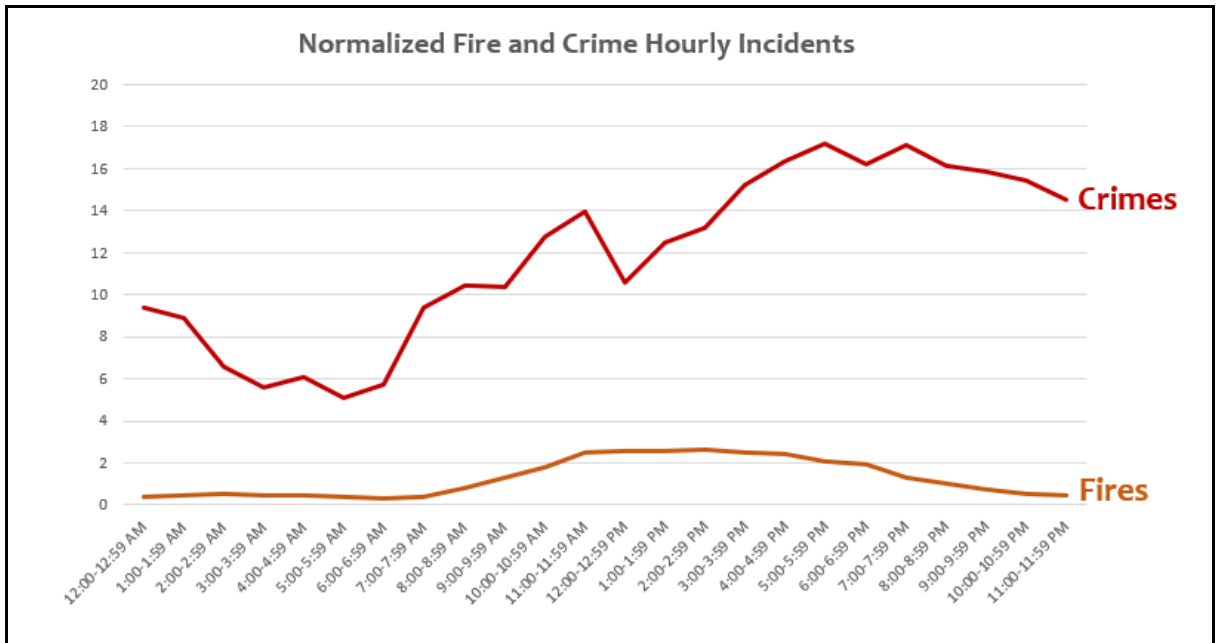

*Figure* 5. Normalized fire and crime-recorded hourly occurrences

Crimes involving property topped the list of crimes (n = 12,432). This was followed by crimes causing harm or intending to cause harm to other people (n = 6,669) and crimes involving controlled substances (n = 4,902). This is partly consistent with the findings of Baculinao and Ceballos (2019), where physical injury (i.e., acts causing harm to another person) is among the top crimes in Central Luzon, where Pampanga is located. When the crime dataset was normalized It was found that crimes are relatively the same in numbers and can be categorized into five classifications (i.e., crimes involving physical injuries, robbery/theft, drugs, reckless imprudence resulting in damage, and petty violations).

## CONCLUSION AND RECOMMENDATIONS

There are several challenges in collecting the data due to the implementation of the Data Privacy Act of 2012 in the Philippines. Despite the mandate of freedom of information in the country for public data. Collection of such data sometimes unnecessarily takes time due to bureaucratic processes and conflicting interpretations of the law and executive orders. Despite all the challenges, the authors of this study were able to collect the two datasets. It was found that the data from the two agencies of the Philippine government were lacking in detailed data. For example, the PNP and BFP data lack basic information on the exact location (e. g., coordinates, blocks, street numbers, etc.), the time interval in their response time, and so on. This missing data could be useful for further optimizing and helping the two agencies in their decision-making processes (Balahadia & Trillanes, 2017; Oñate, 2022).

In the initial analysis, the datasets were able to be presented based on their temporal characteristics. It also showed the leading causes of fire and crime in the province of Pampanga from various timelines. Future research is still needed to further understand how fire and crime occur. Collecting and merging various data from related and relevant agencies concerning fire and crime incidents is also needed to get a better view of what's happening in the province in terms of these incidents. The merged dataset might be useful to support or reject some of the existing studies as well as the results of this study. Finally, the development of relevant systems (e.g., Asor & Sapin, 2020; Balahadia et al., 2020; Villarica et al., 2022) based on the dataset and findings of this study is needed to improve the data collection, analysis, visualization, and policy-making processes for the two agencies (Balahadia & Trillanes, 2017). These recommendations are widely seen as important and should be pursued.

## ACKNOWLEDGEMENT

The authors are forever grateful to the Philippine Bureau of Fire Protection and the Philippine National Police for their support in our research endeavor. We would also like to acknowledge the support provided by Don Honorio Ventura State University through its research management office. This paper is an early version of the paper published in the International Transaction Journal of Engineering, Management, and Applied Sciences and Technologies, volume 13, issue no. 10, pages 1-17. The dataset may be requested from the authors of this study. All information that can identify a person will be omitted when the request is made.

## DECLARATIONS

### *Conflict of Interest*

The authors declare no conflict of interest.

### *Informed Consent*

Not applicable but all requirements to acquire the datasets are registered through eFOI. In addition, the datasets are considered public data.

### *Ethics Approval*

Not applicable. No human or animal subjects were involved in this study.